# Evidence for Kaluza-Klein excitations of the graviton, gluon and quark at LHC

Alain Le Yaouanc[1], François Richard[2]

Université Paris-Saclay, CNRS/IN2P3, IJCLab, 91405 Orsay, France

October 26, 2026

**Abstract**

*Following our paper showing evidence for a sequence of Kaluza-Klein gravitons, we interpret the massive 4 jet excesses observed by CMS in RUN2 and RUN3 data. They could come from the decay of two heavy KK gluons gkk(3.6TeV) and gkk(8.6TeV) into pairs of KK quarks qkk(1TeV) and qkk(2.1TeV). Present limits on gkk into top pairs seem to exclude gkk(3.6TeV) but these limits assume a dominant decay into top pairs, which can be questioned as was the case in our study of KK graviton decays. The cross section predicted for gkk(8.6) requires a BR into a pair of qkk(2.1TeV) or order 10% which is not unlikely. Prospect for reaching a 6 s.d. level local evidence with RUN3 data are promising for 3 channels. KK excitations of the Z boson are also discussed and appear also promising.*

**Contribution to the European strategy for future accelerators**

1 Alain Le Yaouanc <alain.le-yaouanc@ijclab.in2p3.fr>
2 François Richard <francois.richard@ijclab.in2p3.fr>

# Introduction

Up to now the excess [1] observed in RUN2[3] by CMS in the process Y→X+X→JJ+JJ for heavy masses has been interpreted as coming from a pair of squarks, assuming RPV, and from a pair of diquarks [1]. In the present paper we would like to show that it could also fit into a Kaluza Klein interpretation involving a KK gluon decaying into a pair of KK quarks.

In [3-5] evidence for a sequence of KK graviton resonances has been found in the main modes WW/ZZ/hh. A '**smoking gun**' for this interpretation was the disappearance of the indication at 650 GeV in the process VBF→ ZZ when J=0 selections were applied, as expected for a J=2 particle which has an angular distribution similar to the DY background. This effect was interpreted within the Randall Sundrum model predictions for KK gravitons, with however 3 major variations:

- Absence of direct coupling to gluons and photons
- Consequently a dominant production through VBF
- Suppression by a factor 5 to 10 of the BR into top pairs

For high masses the SM background is so low that ATLAS and CMS could use hadronic decays for W/Z, therefore increasing the efficiency for WW/ZZ modes, allowing to observe the KK sequence up to 1.3 TeV, above which there is a continuum of wide resonances.
The mass spectrum of the KK graviton resonances agrees with the predictions of the Randall Sundrum model as recalled in section I.
We have realised that the very heavy KK gravitons can also decay into a pair of lighter KK graviton with significant BR as can be computed from [7]. This opportunity is enhanced when the top pair mode is decreased with respect to the standard RS prediction as observed experimentally. One can have:

$G_{KK}i \to G_{KK}jG_{KK}j$ with $G_{KK}j \to$ WW/ZZ/hh leading to 4 jet topologies for hadronic decays of W/Z/h

If one selects the $G_{KK}i=G_{KK}(1000)$, and sums over all $G_{KK}i$ with masses superior to 2 TeV, this could contribute to the excess observed by CMS for the 4 jet final states XX→JJJJ with mX~1000 GeV. Recall that $G_{KK}(1000)$ is a narrow resonance of width~60 GeV. The expected cross section, as we will see in section I, would however be too small and the width of the 3.5 TeV resonances too large. Figure 1 shows the result from CMS for this channel. In section II we will see that the KK gluon decaying into KK quarks allows to solve this problem.

The KK graviton contribution would of course not explain other observation Y(8600)→X(2100)X(2100) given that the vanishing cross section. One would need to Invoke a gluon KK contribution of the type:

$$g_{KK}(8600) \to g_{KK}(2100)g_{KK}(2100)$$

with a decay of the gluons into two jets. This process is however forbidden by Yang's theorem. as is the case for gg→ $g_{KK}$. One could alternatively have:

$$g_{KK}(8600) \to q_{KK}(2100)\bar{q}_{KK}(2100) \text{ with } q_{KK} \text{ decaying into two jets}$$

Heavy fermionic KK modes decay into electroweak gauge bosons (W, Z, h) plus standard quarks, or cascade decay into lighter KK depending on the mass hierarchy.

CMS has also observed Y(3600)→X(1040)X(1040) which could come from:

$$g_{KK}(3600) \to q_{KK}(1040)\bar{q}_{KK}(1040)$$

3 CMS has partially, ~90 $fb^{-1}$, analysed RUN3 see [1]

One may object that $g_{KK}$(3600) seems excluded by LHC searches in top pairs. These searches however assume that KK gluons decay dominantly into top pairs and from our findings on KK graviton one may question this statement.

The most significant indication claimed by CMS, 3.6 s.d. local, is seen in a non resonant mode X+X with X→JJ with Mx=0.95 GeV. It could come from the sum of the two KK gluon contributions:

$$g_{KK}(3600)+g_{KK}(8600)\rightarrow q_{KK}(1000)\bar{q}_{KK}(1000)$$

# I. Searching for KK gravitons at LHC

## I.1 Graviton decays into SM particles

Let us first recall our findings [3-5] for KK gravitons.:

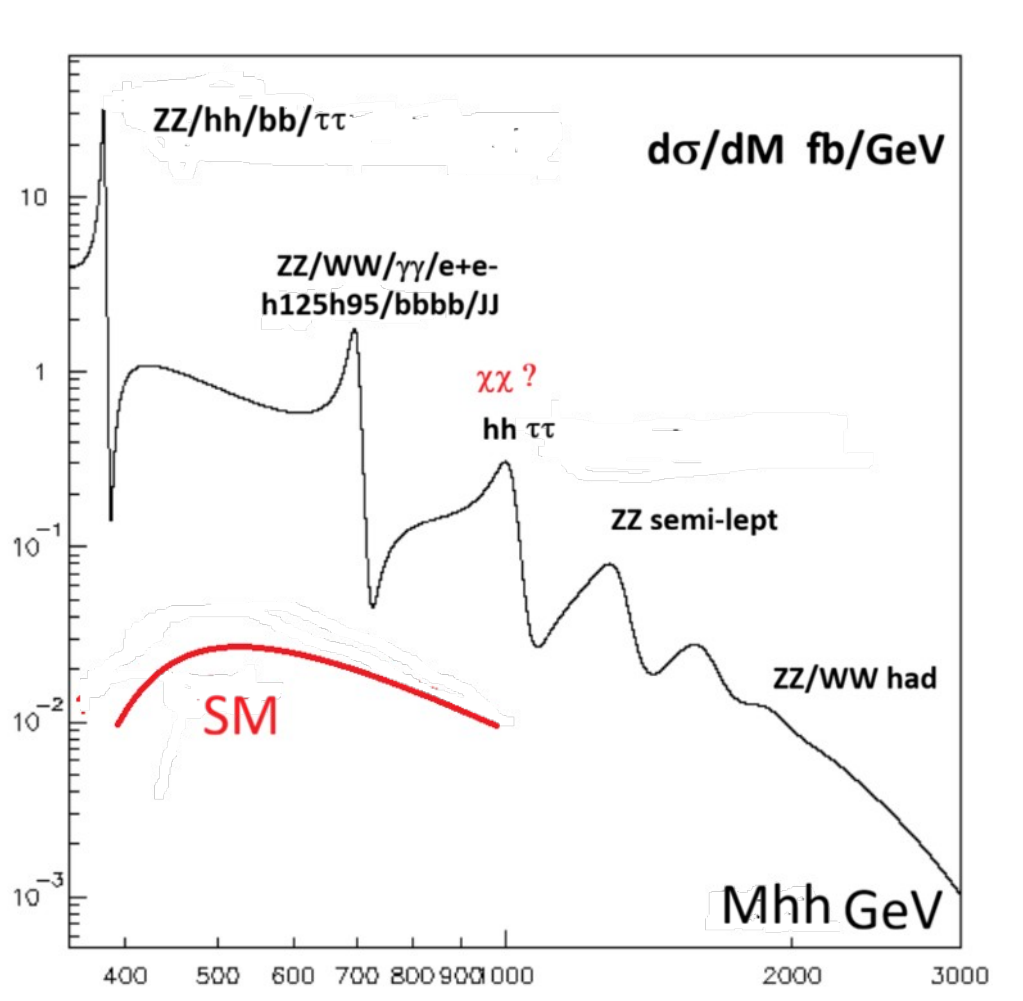

*Figure 1: Predicted VBF differential cross section for KK gravitons→hh. Various observed indications coincident with predicted KK graviton resonances are reported. In red is the SM expectation, an order of magnitude below. The hh component represents ~25% of the BSM cross section.*

**ZZ/WW e+e- γγ hh ττ ZZ semi-leptonic VBF ZZ+WW had**

| m $G_{KK}$ GeV | 380 | **700±20** | 1000 | 1322 | 1640 | 1945 | 2260 | 2590 | 2905 |
|---|---|---|---|---|---|---|---|---|---|
| Γ $G_{KK}$ GeV | 3 | **20±7** | 60 | 130 | 260 | 440 | 680 | 1042 | 1440 |
| VBF→$G_{KK}$ fb | 360 | **100±50** | 38 | 15 | 7 | 3.4 | 1.7 | 0.9 | 0.4 |
| e+e-→ $G_{KK}$ pb | 450 | **130** | 60 | 50 | | | | | |

*Table: Predictions for the KK graviton VBF cross sections and total widths. The coloured line above indicates some observed final states for the various mass regions.*

In the RS model, one predicts a sequence of KK graviton resonances $N_{ikki}$ which have **quantified masses**. These masses are proportional to the **zeros of the Bessel function $J_1$(x):**

**xi=0,3.83,7.02,10.17,13.32,16.5,19.6,22.8,25.9,29.05,32.2,35.3…**

Present indications give **mi~100$x^G_i$,** in GeV**,** while the total width of a spin 2 particle goes like **Γi~0.06($x^G_i$)$^3$** in GeV. The first zero of J1(x), which corresponds to the standard graviton gives, as one would expect, m=0.

So far we have candidates at 380 GeV (5), 700 GeV (9) , 1 TeV (2), 1.3 TeV(1) and a continuum of wide resonances observed in hadronic modes from WW/ZZ/hh for higher masses up to 3 TeV [4].

The large hh contribution shown in figure 1 dominates over the SM Higgs self-coupling contribution h*→hh. The observed excess seems milder than expected which we interpret as inadequate selections for a spin 2 resonance.

## II.2 The process $G_{KKi} \to G_{KK}380G_{KK}380$

The process $G_{KKi} \to G_{KK}380G_{KK}380$ computed in [7], would provide the ultimate proof of the existence of heavy KK gravitons and constitute an important test of the RS model. These processes become dominant above 2 TeV. The table below shows that there are significant contributions of this mechanism which could be identified with more integrated luminosity.

| m $G_{KK}$i GeV | 1322 | 1640 | 1945 | 2260 | 2590 | 2900 |
|---|---|---|---|---|---|---|
| VBF fb | 15 | 7 | 3.4 | 1.7 | 0.9 | 0.4 |
| Total width GeV | 130 | 260 | 550 | 1460 | 2900 | 5600 |
| BR(G380G380) % | 2 | 12 | 20 | 53 | 63 | 75 |
| $G_{KK}$i → G380380 fb | 0.3 | 0.8 | 0.7 | 0.9 | 0.6 | 0.3 |

Decay modes are WWWW 25%, ZZWW 25%, hhWW 25%, hhZZ 12.5% and hhhh+ZZZZ 12.5%. A detailed study has to be performed. One should try to reconstruct WWWW events in purely hadronic (20%) and semi-leptonic (80%) modes which should provide a high detection efficiency and sufficient purity given the high masses involved.

# II. Searching for $g_{KKi} \to q_{KKj}\bar{q}_{KKj}$ at LHC

## II.1 Results

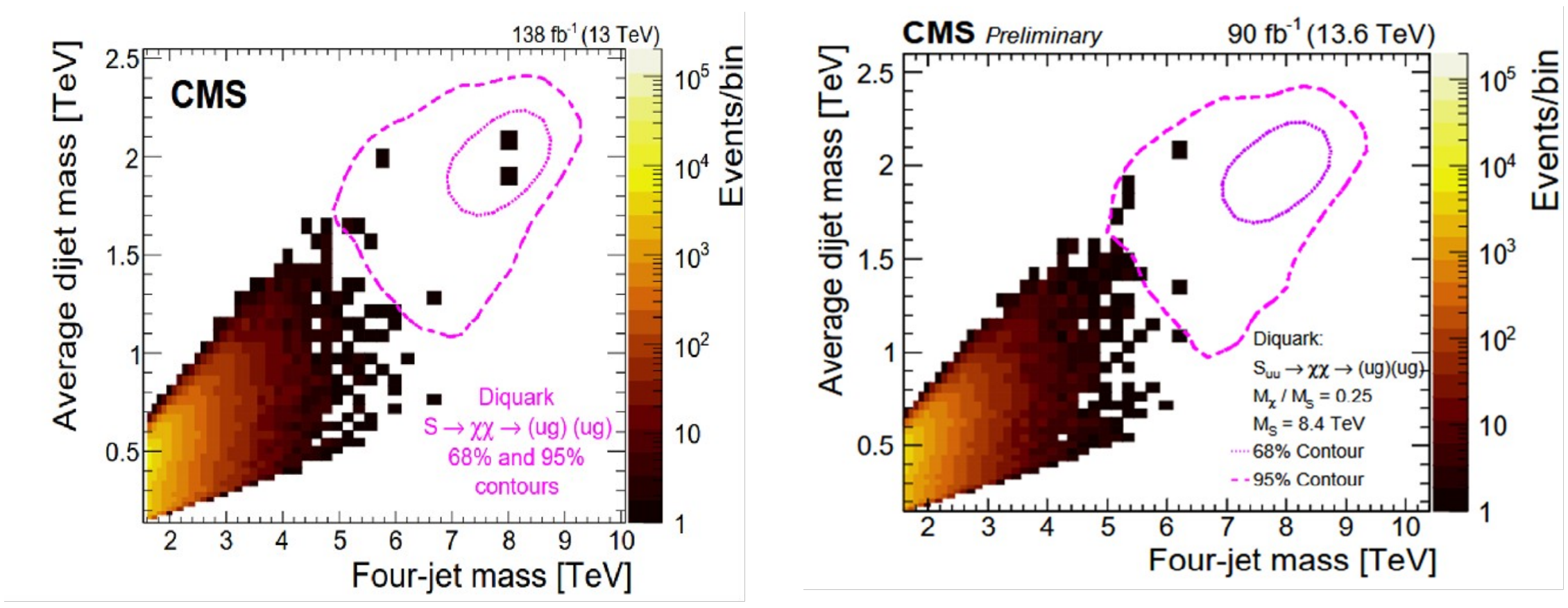

*Figure 2: Search for 4 jet Y→X+X with X->JJ events at 13 TeV and 13.6 TeV.*

Figure 2 shows two striking near-by events, with four jets at 8 TeV, giving two jets with an average mass of 2 TeV, which are almost background free. The proposed interpretations predict resonances with a wide spread, as indicated by the magenta contours. A partial analysis of RUN3 data did not bring additional 'gold plated' events.
ATLAS [2] has produced a similar analysis, without observing 'gold plated' candidates as shown in figure 3 where we have drawn a similar contour to figure 2.

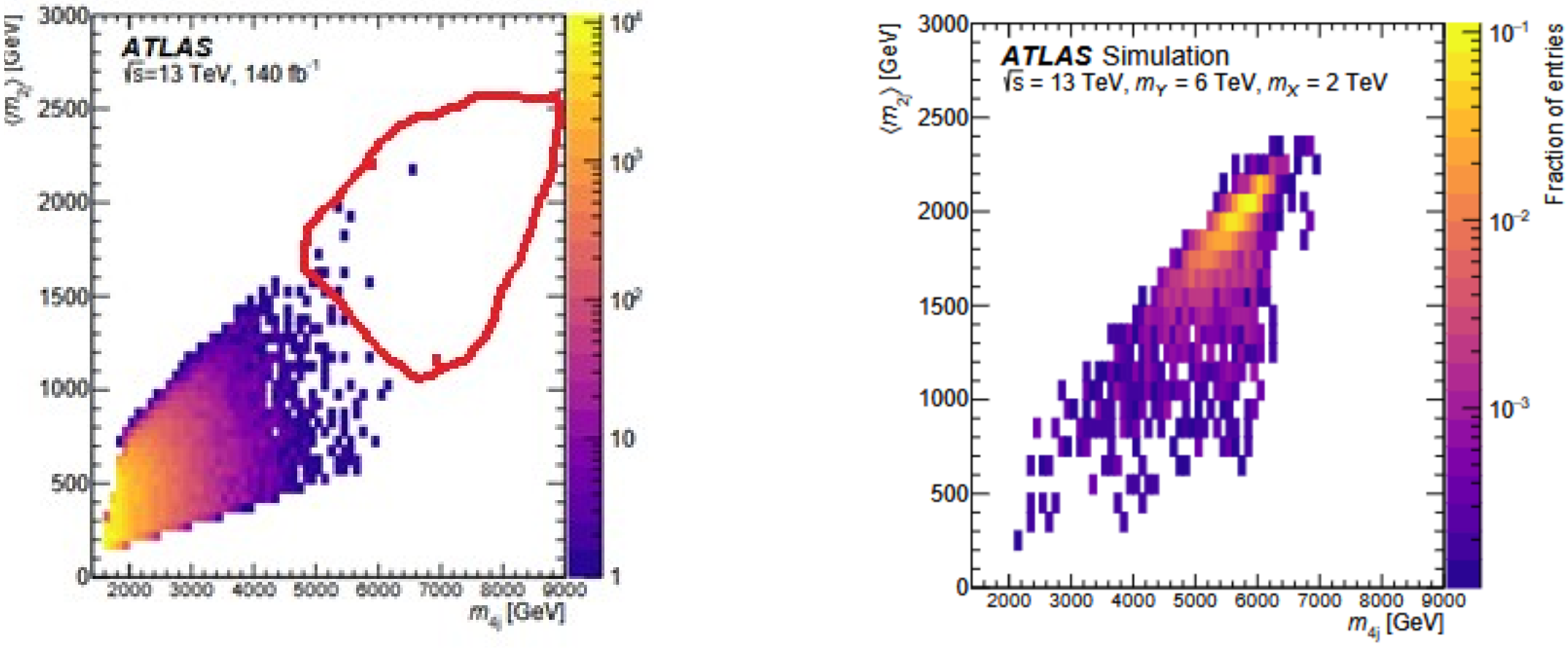

*Figure 3: Results from ATLAS with a red contour similar to figure 2. On the right side is the simulated background.*

The significances of the two heavy resonances claimed by CMS is shown in figure 4.
They depend on the assumption of the ratio of MX/MY, in our interpretation on the ratio of the KK quark mass to the KK gluon mass and on the width assumption. The  mass ratios need not be equal. The highest significance for the 8.6 TEV candidate corresponds to KK quark mass of 2.15 TeV, while for the 3.6 TeV one has a quark mass of 1.04 TeV, favouring a width of 360 GeV. Both reach a local significance close to 4 s.d.

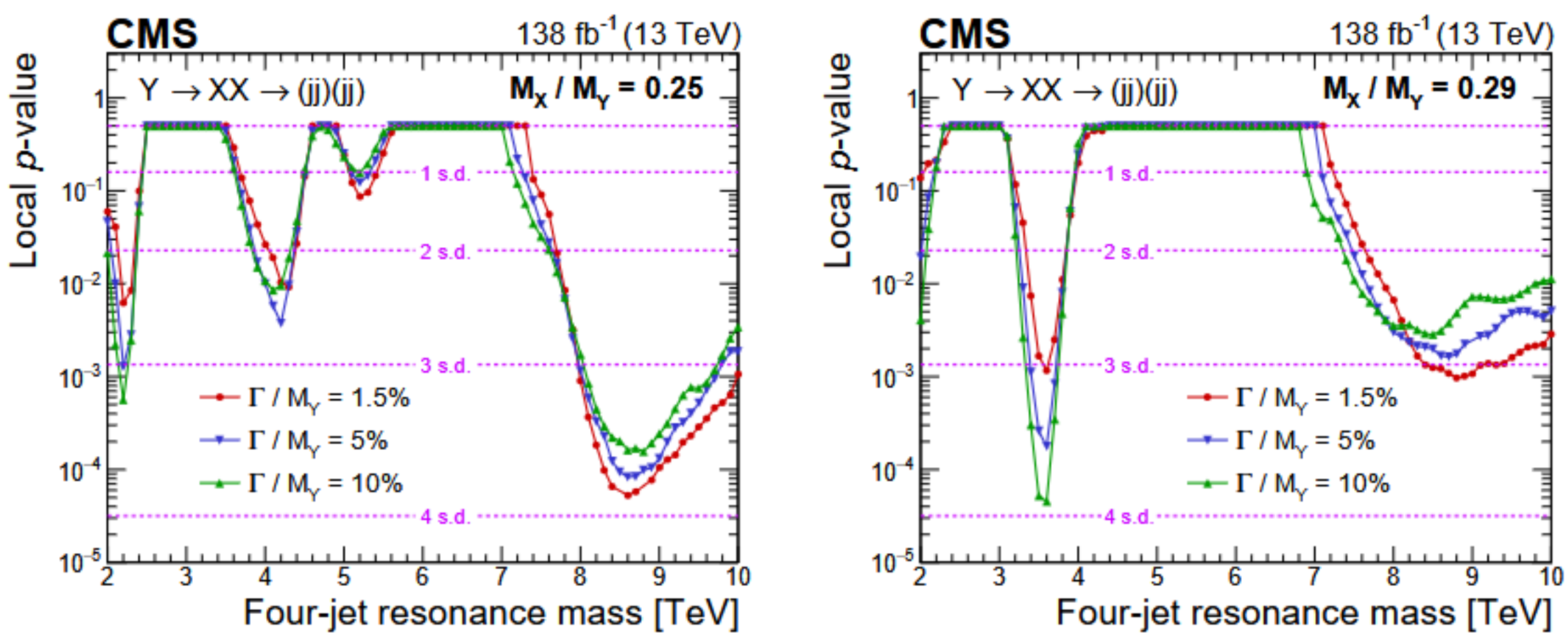

Figure 4:Observed local p-value for a four-jet resonance Y, decaying to a pair of dijet resonances, X, with MX/MY=0.25 (left) and 0.29 (right), and various widths of Y superimposed. Also shown are corresponding levels of local significance (dashed) in s.d.

The other evidence is in XX, leaving free Y as shown in figure 5. We interpret this as saying that this channel receives a contribution from both KK gluons[4]. CMS claims that it has the highest significance which is not quite true in our case where we allow the ratio X/Y to vary.

4 On this plot which mixes both KK gluon resonances, the 2.1 GeV resonance originating from the 8.6 TeV resonance almost disappears being drowned by the huge QCD background.

So far RUN2 data have not been combined with RUN3 data.

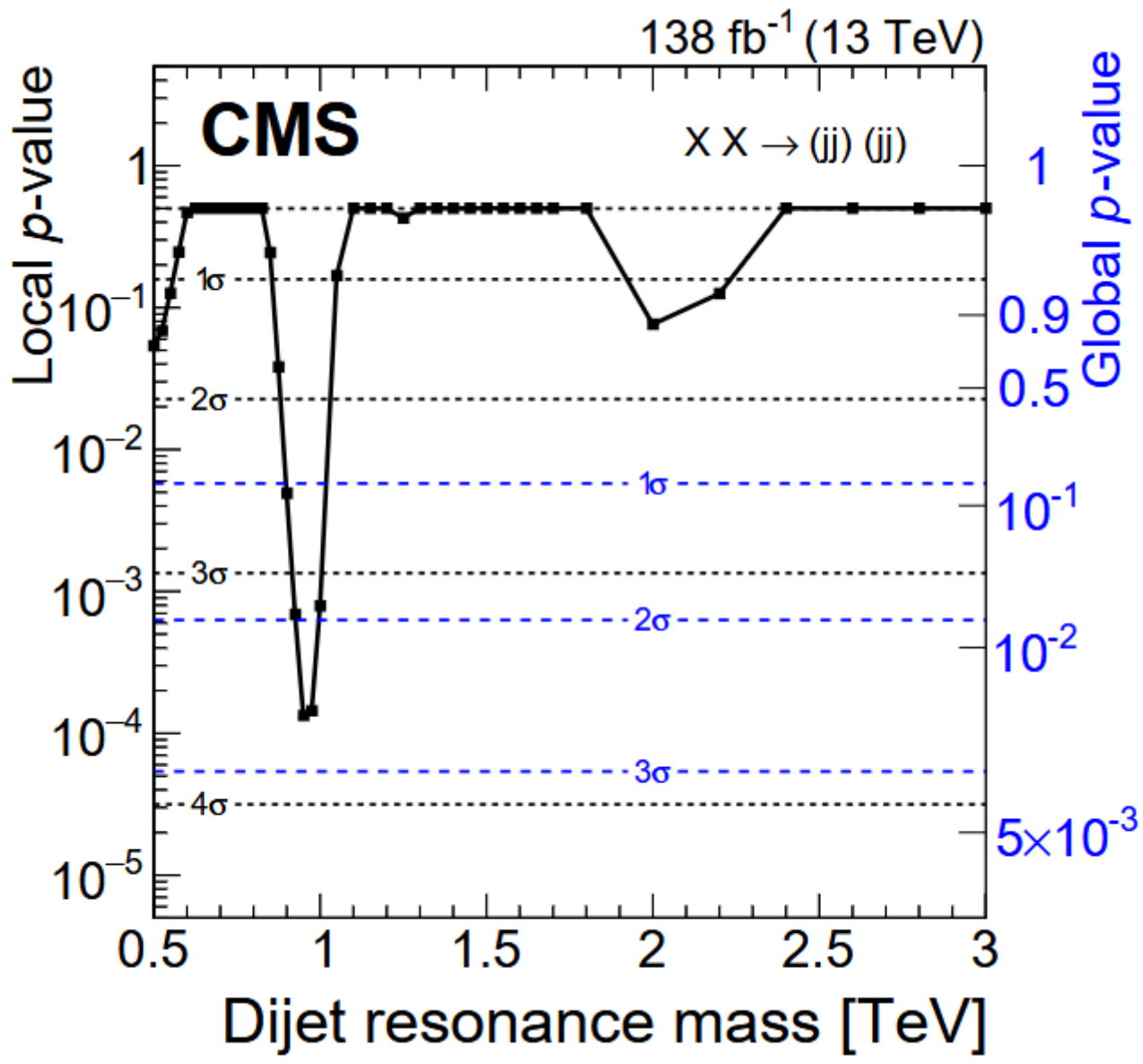

*Figure 5: Observed p-values (points) for nonresonant production of a paired dijet resonances, X, Also shown are corresponding levels of local significance (dashed) and global significance (blue dashed) in units of standard deviation.*

## II.2 Comparison to the RS predictions

For $g_{KK} \to q_{KK} \bar{q}_{KK}$ we expect a BR roughly comparable to the process $g_{KK} \to t \bar{t}$ with a mild s-wave suppression which is negligible for the 8.6 TeV case and rather small for 3.6 TeV . Given that in the RS model what counts is the localisation of the top and the KK fermions in the extra-dimension, which are a priori unknown, the exact prediction seems out of reach.
The production cross section is given by figure 6. It is equal to 1.4 fb for the 8.6 TeV gKK and 110 fb for the 3.6 TeV resonance. The significance of the deviations are similar reflecting that the SM background is about four orders larger at 3.6 TeV.

The decay width in figure 6 assumes a dominant decay of KK gluons into top pairs. To understand why the LHC mass limits on $g_{KK}$ are optimistic one needs to assume a reduced coupling to top pairs as was the case for the KK gravitons which, in turn, means a reduced total width. At 3.6 TeV the measured width is of order 200 GeV while figure 2 gives 700 GeV, which confirms our expectation of a reduction of the top pair BR. The width of a vector particle grows linearly with mass implying a width of 500 GeV at 8.6 TeV. Figure 2 seems to indicate a larger width but this could be due to the effect of rising luminosity for low masses.
In the RS model, KK quarks decay into the same flavour quark and a SM boson (photon, gluon, Z, W, h). These states, except for the photon, decay mostly into hadrons and therefore heavy resonances give two jets, agreeable to the CMS analysis.
In RUN2, CMS has observed 7 events within the defined contours, meaning:

$$BR=7/(140fb^{-1}*1.4fb*A)=3.5\%/A, \text{ where A is the acceptance}$$

We need a BR ~ 10%, that is A~30%.

At 3.6 GeV this cross section, given in figure 6, is equal to 110 fb that is 80 times higher. Given that the significance is similar this means that the background is 6000 times higher, which matches observation.

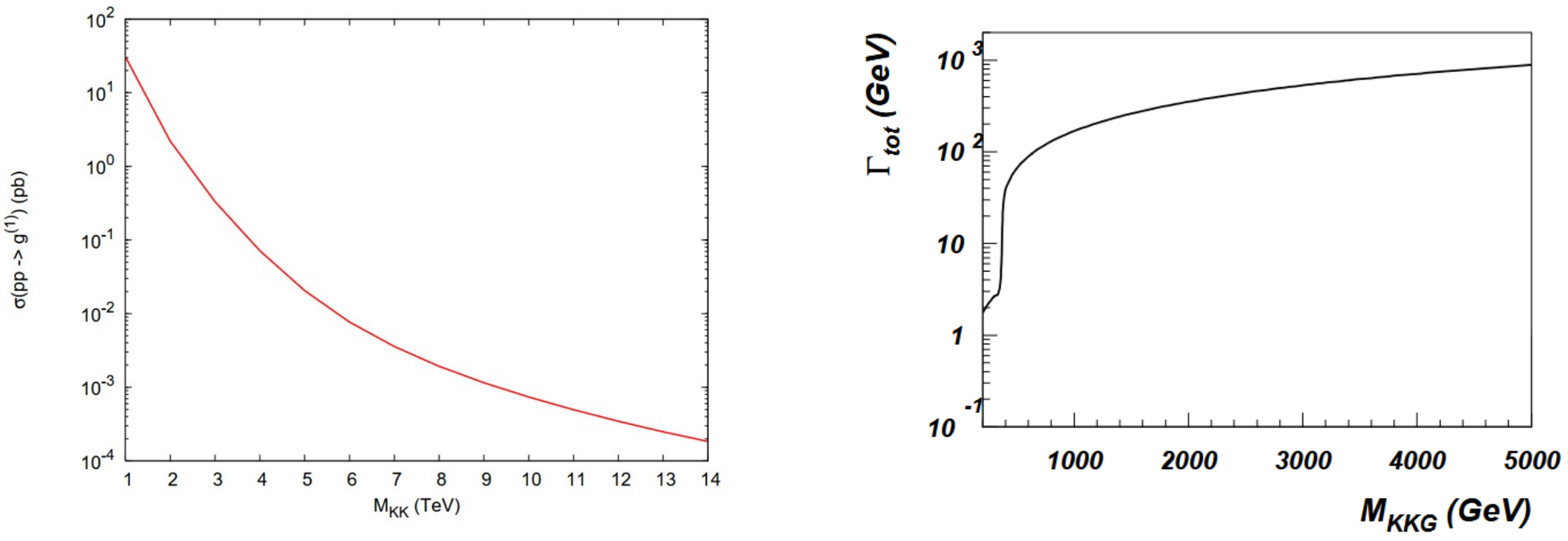

*Figure 6: Predicted [10] cross sections and widths for a KK gluon in the RS model at LHC.*

The sequence of masses is well reproduced by the RS prediction of masses proportional to 2.45, 5.57, 8.7, 11.84 for KK vectors [6]. The detailed calculation of these ratios can be found in [14]. One predicts 5.57/2.45=2.27 while CMS gives 8600/3600-4000=2.38-2.15. The next KK gluon would be at 13 TeV well beyond LHC reach.
For the KK quarks one predicts [8] a factor 2 ratio while one measures 2.15/1.04, in fair agreement. The **third KK quark is predicted at 3.1 TeV**, therefore accessible in the decays of $g_{KK}$(8.6) into KK quark pairs.

Experimentally one finds: Graviton/Quark/Gluon=380/1000/3600 = 1/3/10. $Z_{KK}$ seems to have a mass comparable to KK gluons as predicted by the Zbb measurement at LEP [11] discussed in the next section.

This apparent success of the RS predictions for the sequence of masses hides a serious problem, which is the mass ratio between the KK gluon and the KK graviton masses. In the naive model [6] one has graviton1/gluon1=3.83/2.45=1.56 while experiment suggest 380/3600~0.1 !
References [9] and [17] suggest that this can be interpreted by assuming a **breakdown of universality of KK mass scales** induced by **brane kinetic terms** acting differently on gravitons and gluons.

# III. Searching for $Z_{KK}$

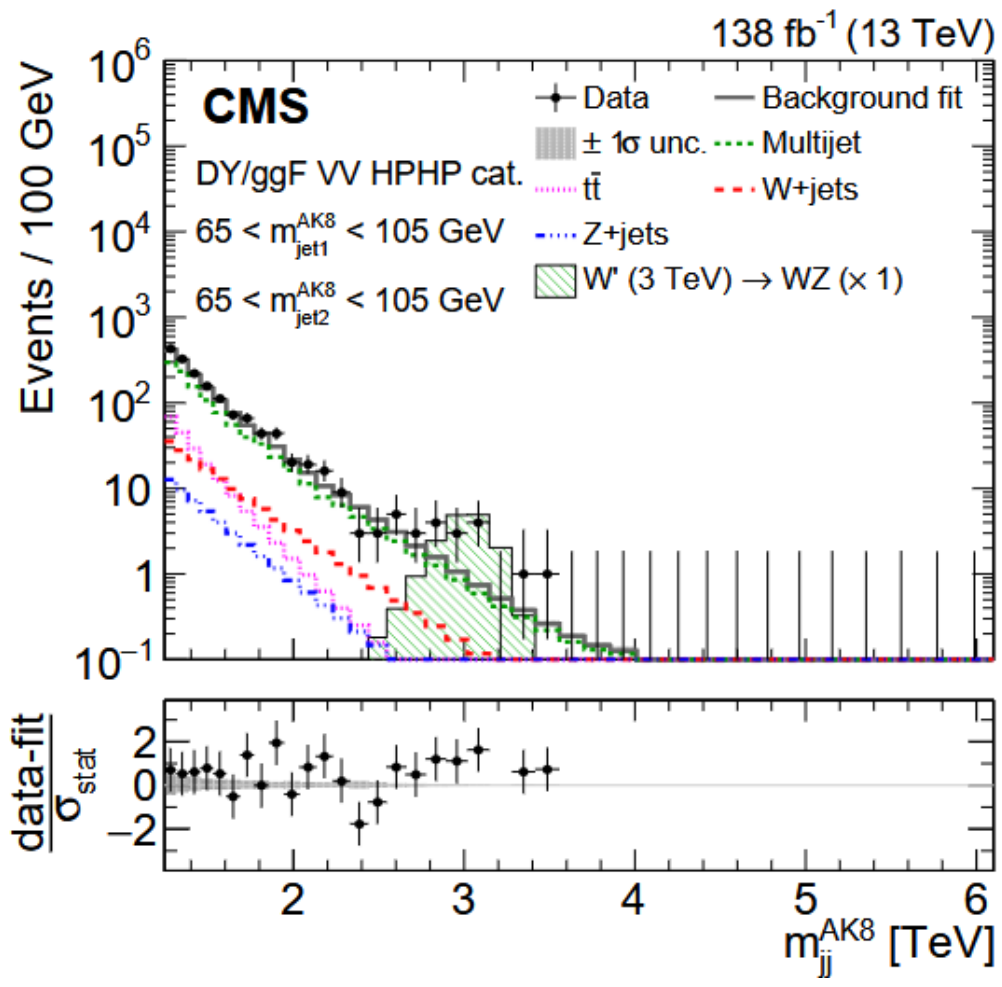

+

*Fig 7: Mass distribution of the VV candidates decaying hadronically with a visible excess around 3 TeV.*

KK recurrences of the Z are also produced by quark anti-quark annihilation but with a cross section two orders of magnitude below the KK gluon recurrence. One also predicts that it should be searched into top pairs. If one assumes, as for KK gravitons, that this is not so, the main decay could be through vector bosons

WW and ZH (ZZ is forbidden by Yang's theorem). If one also assumes that the $Z_{KK}$ is very heavy as suggested by precision measurements at LEP1 [11], one can use hadronic decays as for the heavy KK gravitons [8]. One observes an indication at 3 TeV GeV, as shown in figure 7. The next KK should be at 7 TeV, out of reach at LHC given its minute cross section.

To satisfy precision measurements, such a low mass for $Z_{KK}$ requires an extended symmetry with the presence of a Z' with similar mass, therefore also detectable at LHC. In reference [4] we have shown that one expects [5]:

$$\mathbf{mZ_{KK}=2.5TeV/sin\theta'}$$

where $\theta'$ is a mixing angle which fixes the parameter of the theory, allowing to predict the $q\bar{q}\rightarrow Z_{KK}$ cross section. The solution $mZ_{KK}$=3 TeV gives **sinθ'=0.83.**

One predicts that $mZ_{KK}$~mZ', therefore what LHC measures is the sum of the two resonance contributions with, presumably, contributions from $W_{KK}$. The predicted total width at 3 TeV is ~100 GeV.

The excess in figure 7 corresponds to ~10 events. Note that the mass cut <100 GeV partially eliminates Zh and Wh contributions. Reference [14] predicts a 1 fb cross section at 3 TeV for Z'→WW, where Z' means the sum of $Z_{KK1}$ and Z' contributions. Assuming a 15% reconstruction efficiency for the WW hadronic modes, one expects 10 events, in agreement given this crude evaluation.

Including properly Zh and Wh should double the efficiency, therefore worth trying.

The predicted Z' cross section into lepton pairs is of order 0.01 fb for the e+e- + μ+μ- final states. Figure 8 shows that CMS observes a mild excess at 3 TeV. The corresponding cross section, 0.05±0.02fb, is barely compatible with the prediction, meaning that Z' couples more strongly than expected. This feature was noticed for the $G_{KK}$700 graviton coupling to e+e- indicated by ATLAS and CMS, which was exceeding predictions.

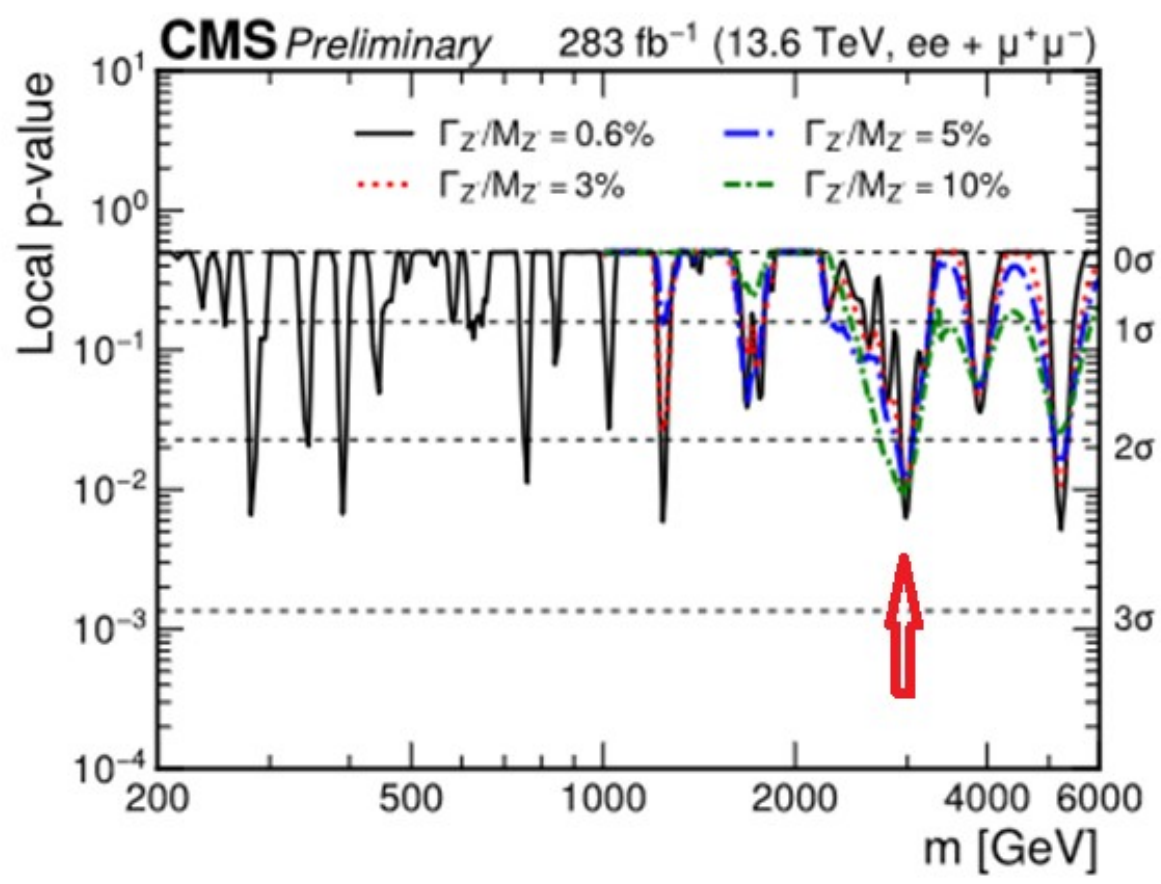

*Figure 8: Observed p-value by CMS for the di-lepton spectrum based on a luminosity of 283 fb-1.*

5This solution from [4] differs from [11] to allow an increased coupling of KK gravitons to e+e-

# Conclusion

We have provided a plausible explanation for various excesses observed by CMS in four jet final states. This interpretation follows closely the RS model with the exception of the BR into top pairs of the KK gluon resonance. The present local significance for the four evidences, 8.6 TeV and 3.6 TeV for KK gluons, 2.1 TeV and 1 TeV for KK quarks, comes close to a local 4 s.d. significance. **The ratio of masses for KK gluons and KK quarks** for the **first two generations** follows the **RS predictions**. There is however a serious issue in understanding the ratio between the KK graviton and gluon masses which strongly differ from the naive model.

We predict that one could also observe $g_{KK}$(8.6)→ $q_{KK}$(3.1)+$q_{KK}$(3.1), a spectacular final state, **background free**, which would constitute a crucial confirmation of the model.

Precision measurements from LEP1 indicate that a $Z_{KK}$ could also be soon discovered at LHC and one sees a first indication for this in WW+ZH+WH hadronic modes at 3 TeV and in e+e- + μ+μ- at the same mass, agreeing with predictions. If confirmed, it implies the presence of an **additional Z'** resonance with a similar mass. Such a scenario can also be indirectly tested by precision measurements in an e+e- collider [13].

Adding all CMS data from RUN3 data should allow to push the significance of the 4 jet events to more than 6 s.d. Given that there are 3 channels and given the potential represented by ATLAS data, LHC could soon reach a fundamental discovery. Admittedly the partial analysis of RUN3 by CMS and results from ATLAS on RUN2 did not bring additional 'gold plated' four jet events.

Evidences for $Z_{KK}$+Z'+W'(3TeV)→WW +ZW and $Z_{KK}$+Z'(3TeV)→e+e- + μ+μ- give weaker but consistent indications for $Z_{KK}$ +Z'+W' and need to be confirmed.

**Acknowledgements**

*The wish to thank Caroline Collard for providing the useful contacts within the CMS collaboration.*
*Thanks to Matt LeBlanc for pointing out the 4 jet analysis from ATLAS.*